\documentclass[twocolumn,tightenlines,showpacs,prl, floatfix,
preprintnumbers,amsmath,amssymb]{revtex4}

\usepackage{graphicx}

\usepackage{bm}

\begin{document}

\title{Gravitational wave stochastic background from cosmic (super)strings}

\author{Xavier Siemens$^1$, Vuk Mandic$^2$ and Jolien Creighton$^1$}
\affiliation{
$^1$Center for Gravitation and Cosmology, Department of Physics, 
University of Wisconsin --- Milwaukee, P.O. Box 413, Wisconsin, 53201, USA\\
$^2$LIGO Laboratory, California Institute of Technology, MS 18-34,
Pasadena, CA 91125, USA\\
}

\date{\today}

\begin{abstract} 
  We consider the stochastic background of gravitational waves
  produced by a network of cosmic strings and assess their
  accessibility to current and planned gravitational wave detectors,
  as well as to big bang nucleosynthesis (BBN), cosmic microwave
  background (CMB), and pulsar timing constraints. We find that
  current data from interferometric gravitational wave detectors, such
  as LIGO, are sensitive to areas of parameter space of cosmic string
  models complementary to those accessible to pulsar, BBN, and CMB
  bounds.  Future more sensitive LIGO runs and interferometers such as
  Advanced LIGO and LISA will be able to explore substantial parts of
  the parameter space.
\end{abstract}


\maketitle

\noindent \textit{I. Introduction.} -- Cosmic strings can be formed in
phase transitions in the early universe \cite{kibble76}, and are
viable candidates for generating a host of interesting astrophysical
phenomena \cite{alexbook}. Cosmic superstrings are produced in certain
string-theory inspired inflation scenarios \cite{cstrings}.  Since
fundamental strings interact probabilistically, and due to the higher
dimensionality of string theories \cite{cstrings}, cosmic superstrings
re-connect with a probability $p$ that can be smaller than unity
($p=1$ for field theoretic strings).  Values of $p$ are expected to
lie in the range $10^{-3}-1$ \cite{jjp}.  In stringy scenarios it is
also possible to form more than one kind of string.  Here we assume
that only one kind of string forms, and a network density proportional
to $p^{-1}$.
 
Cosmic (super)strings can produce strong bursts of gravitational
radiation.  This possibility was first considered by Berezinsky,
Hnatyk and Vilenkin \cite{firstcusps}, and later explored in detail by
Damour and Vilenkin \cite{DV0,DV2}.  The strongest bursts are produced
at cosmic string cusps (regions of string that acquire large Lorentz
boosts) and could be detected even by Initial LIGO
\cite{DV0,DV2,SCMMCR}. The gravitational waveforms of cusps are simple
and robust to classical perturbations \cite{kenandi} as well as
quantum effects \cite{damournew}.

\begin{figure}[hbtp]
\includegraphics[width=3.3in]{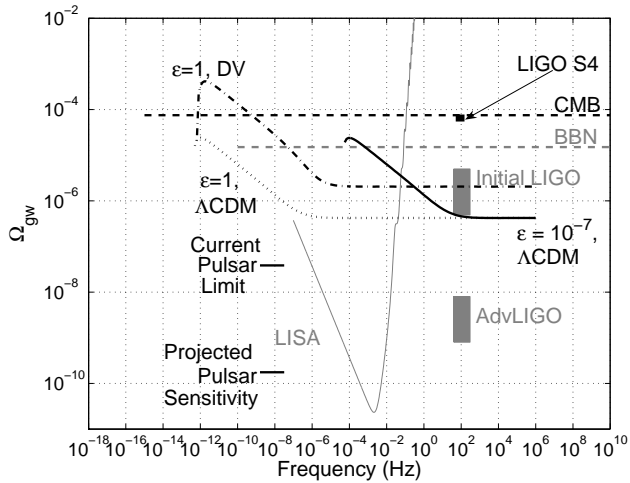}
\caption{Different experimental bounds and future experimental
  sensitivities are shown in relation to the cosmic string spectra
  computed for $p=5\times 10^{-3}$, $G\mu = 10^{-7}$.  The dot-dashed
  curve was computed for $\varepsilon = 1$ using Eqs.~(4.1-4.7) of
  Damour \& Vilenkin \cite{DV2}.  The solid and dotted curves were
  computed using the method described in this paper and $\varepsilon =
  10^{-7}$ and 1 respectively.  Note that the model depicted by the
  solid curve is not accessible to the pulsar or BBN bounds, but may be
  accessible to Initial LIGO \cite{ILIGO}.  The BBN and CMB bounds
  apply to the integral of the spectrum over the frequency range
  indicated by the corresponding lines. See text for more detail.}
\label{landscape}
\end{figure}

Cosmic (super)strings also produce a stochastic background of
gravitational waves (GWs) \cite{alexbook,caldwellallen,DV0,DV2}, whose
spectrum is usually defined as $\Omega_{\rm gw}(f) = (f/\rho_c) \;
d\rho_{\rm gw}/df$.  Here, $d\rho_{\rm gw}$ is the energy density of
GWs in the frequency range $f$ to $f+df$ and $\rho_c$ is the critical
energy density of the Universe. We examine the GW background produced by
the incoherent superposition of cusp bursts from a network of cosmic
strings. We build on the results of Damour and Vilenkin \cite{DV0,DV2}
generalising them in two ways: 1) we consider a generic cosmological
model, that allows us to include the effects of late time acceleration
(see \cite{SCMMCR}), and 2) we generalise the analysis to include
arbitrary loop distributions.  The former generalisation results in a
stochastic background within an order of magnitude of, but smaller
than, the estimates of \cite{DV0,DV2} (see Fig.~\ref{landscape}). The
latter generalisation allows us to compute the background when string
loops are large when they are formed and thus long-lived, a
possibility suggested by recent numerical simulations
\cite{recentsims1,recentsims2}.  Recently, Hogan \cite{hogan} has made
analytic estimates for the case when the size of loops at formation is
about a tenth of the horizon.

We investigate the detectability of the background by a wide range of
experiments. We consider the LIGO bound from the fourth science run S4
(Bayesian 90\% upper limit $\Omega_{\rm gw} < 6.5 \times 10^{-5}$ in
51-150 Hz band \cite{S4stoch}), the bound based on pulsar timing
experiments (95\% detection rate upper bound $\Omega_{\rm gw} < 3.9
\times 10^{-8}$ at frequencies $1/(20 {\rm yr}) - 1/{\rm yr}$
\cite{pulsar}), as well as the expected future reaches of LIGO,
Advanced LIGO, LISA \cite{LISA}, and pulsar timing experiments
\cite{pulsar}.  We also consider the indirect bound due to big-bang
nucleosynthesis (BBN) \cite{maggiorereview}: $\int \Omega_{\rm gw}(f)
d(\ln f) < 1.5 \times 10^{-5}$, assuming 4.4 as the 95\% upper limit
on the effective number of neutrino species at the time of BBN
\cite{cyburt}. This bound applies to the signal produced before the
time of BBN, i.e. to redshifts $z > 5.5 \times 10^9$, and to
frequencies above $\sim 10^{-10}$ Hz (corresponding to the comoving
horizon size at the time of BBN).  Similarly, we consider the bound
obtained using the CMB and matter spectra \cite{smith} $\int
\Omega_{\rm gw}(f) d(\ln f) < 7.5 \times 10^{-5}$ (95\% confidence
limit, assuming adiabatic initial conditions). This bound applies to
signals produced before photon decoupling, i.e. for $z > 1100$, and to
frequencies above $\sim 10^{-15}$ Hz (corresponding to the comoving
horizon size at the time of photon decoupling).  Finally, we consider
the projected sensitivity of the LIGO burst search, optimized to
search for individual cusp bursts at relatively low redshifts
\cite{SCMMCR}.  For the above limits on $\Omega_{\rm gw}$, as well as
in the remainder of the paper, we assume a value of the Hubble
parameter $H_0 = 73$ km/s/Mpc \cite{hubble}. The $5\%$ uncertainty in
the value of the Hubble parameter does not alter our conclusions.
Figure \ref{landscape} shows the different experimental bounds in
relation to examples of the cosmic string spectrum. We will show that
these experiments explore a large fraction of the cosmic string
parameter space, making burst and stochastic GW searches rare and
powerful probes of early universe physics and string theory motivated
cosmology.

\noindent \textit{II. The stochastic background.} -- We have used the
results of Allen and Romano \cite{allenandromano}, to compute the GW
spectrum by evaluating the strain at a point in space \cite{SMC},
\begin{equation}
\Omega_{\rm gw}(f) = \frac{4 \pi^2}{3H^2_0}f^3 
\int dz \int dl \, h^2(f,z,l)  \frac {dR}{dzdl}.
\label{e:Omega(f)2}
\end{equation}
In the following, we describe the quantities that enter this
expression. The strain produced by a cusp at a redshift $z$, from a
loop of length $l$, can be read off Eq.~(46) of \cite{SCMMCR}:
\begin{equation}
h(f,z,l)  = g_1 \frac{G\mu l^{2/3} H_0} {f^{4/3}(1+z)^{1/3}  
\varphi_r (z)}.
\label{eq:hfz1}
\end{equation}
Here $g_1$ absorbs the uncertainty on the amount of length $l$
involved in the production of the cusp \cite{SCMMCR}, $G$ is Newton's
constant, and $\mu$ is the mass per unit length of strings. We expect
the ignorance constant $g_1$ to be of $O(1)$ provided loops are
smooth.  The dimensionless function $\varphi_r (z)$ relates the proper
distance to the redshift (see Appendix A of \cite{SCMMCR}). The burst
rate entering Eq.~(\ref{e:Omega(f)2}) is given by Eq.~(58) of
\cite{SCMMCR}:
\begin{equation}
\frac{dR}{dzdl} = H^{-3}_0 \varphi_V(z) (1+z)^{-1} \nu(l,z)
\Delta(f,z,l).
\label{eq:rate11}
\end{equation}
Here $\varphi_V(z)$ is a dimensionless function that relates the
volume element to the redshift (see Appendix A of \cite{SCMMCR}) and
the factor $(1+z)^{-1}$ comes from the relation between the observed
burst rate and the cosmic time. The number of cusps per unit
space-time volume from loops with lengths in the interval $dl$ at a
redshift $z$ is $\nu(l,z)dl = (2c/l) \; n(l,z)dl$ (see \cite{SCMMCR}).
Here, $c$ is the number of cusps per loop oscillation (assumed to be 1
in the analysis below), and $n(l,z)$ is the loop distribution which we
vary in the analysis below.  The fraction of bursts we can observe is
$\Delta(f,z,l) \approx \theta_m^2(z,f,l) \Theta(1-\theta_m(z,f,l))/4$,
with $\theta_m (z,f,l)= [g_2 (1+z)f l]^{-1/3}$. The ignorance constant
$g_2$ absorbs factors of $O(1)$, as well as the fraction of the loop
length $l$ that contributes to the cusp \cite{SCMMCR}. We expect $g_2$
to be of $O(1)$ if loops are smooth. The angle $\theta_m$ is the
maximum angle that the line of sight and the direction of a cusp can
subtend and still be observed at a frequency $f$. Thus $\theta_m^2/4$
is the beaming fraction corresponding to the angle $\theta_m$, and the
$\Theta$ function cuts off events that don't have the form of
Eq.~(\ref{eq:hfz1}).

\begin{figure*}[hbtp]
$\begin{array}{cc}
\includegraphics[width=2.8in]{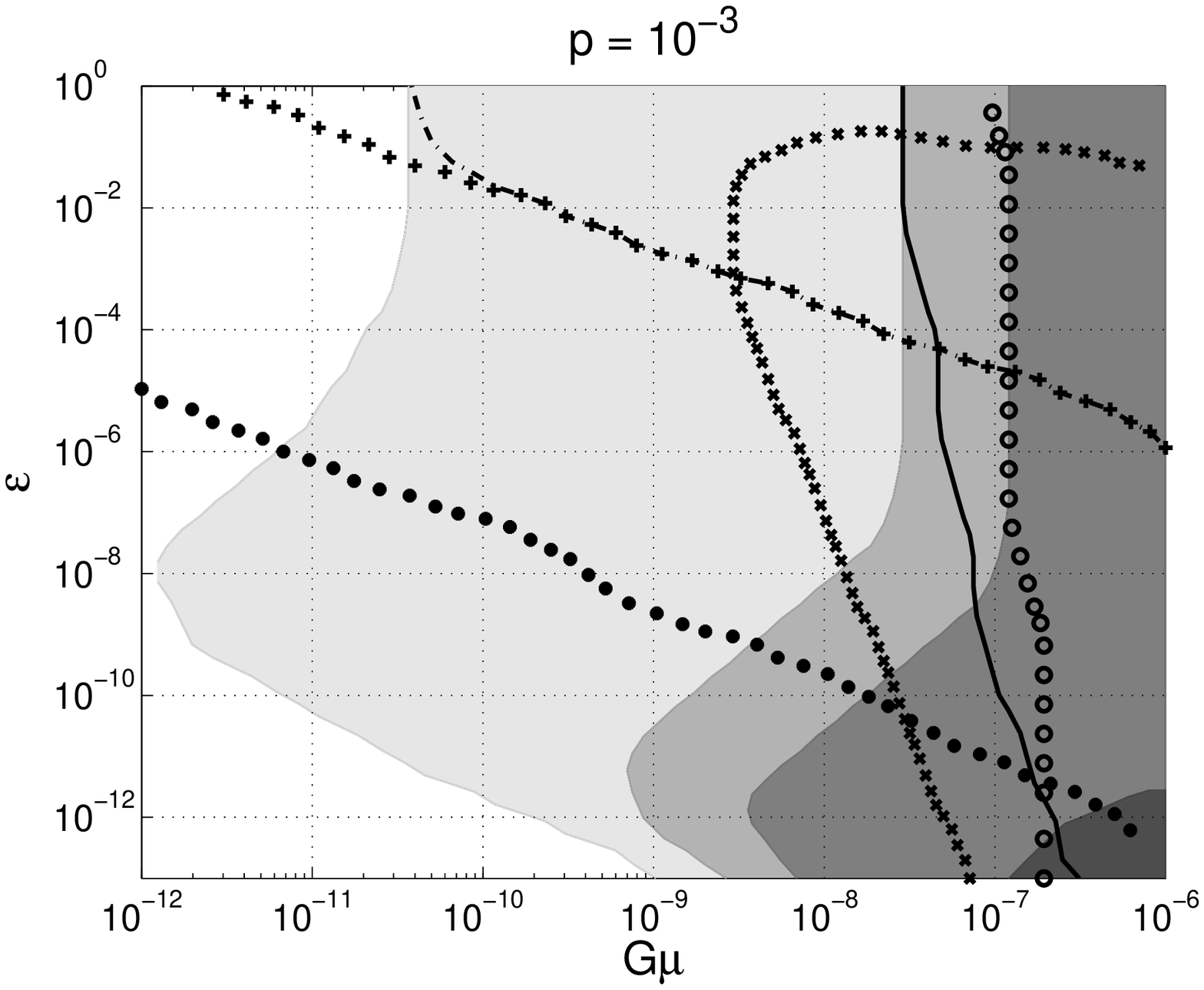} &
\hspace{1cm}
\includegraphics[width=2.8in]{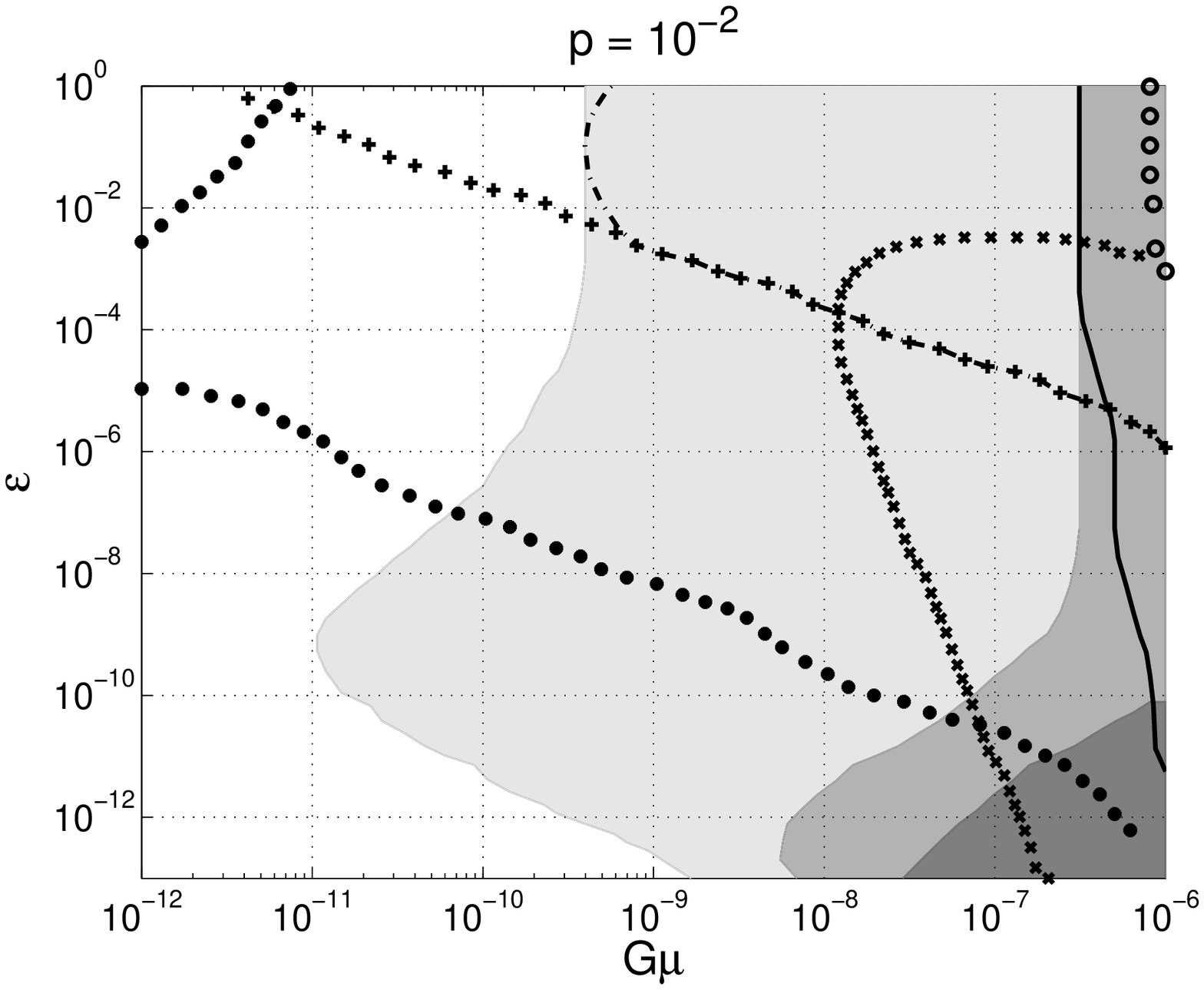} \\
\includegraphics[width=2.8in]{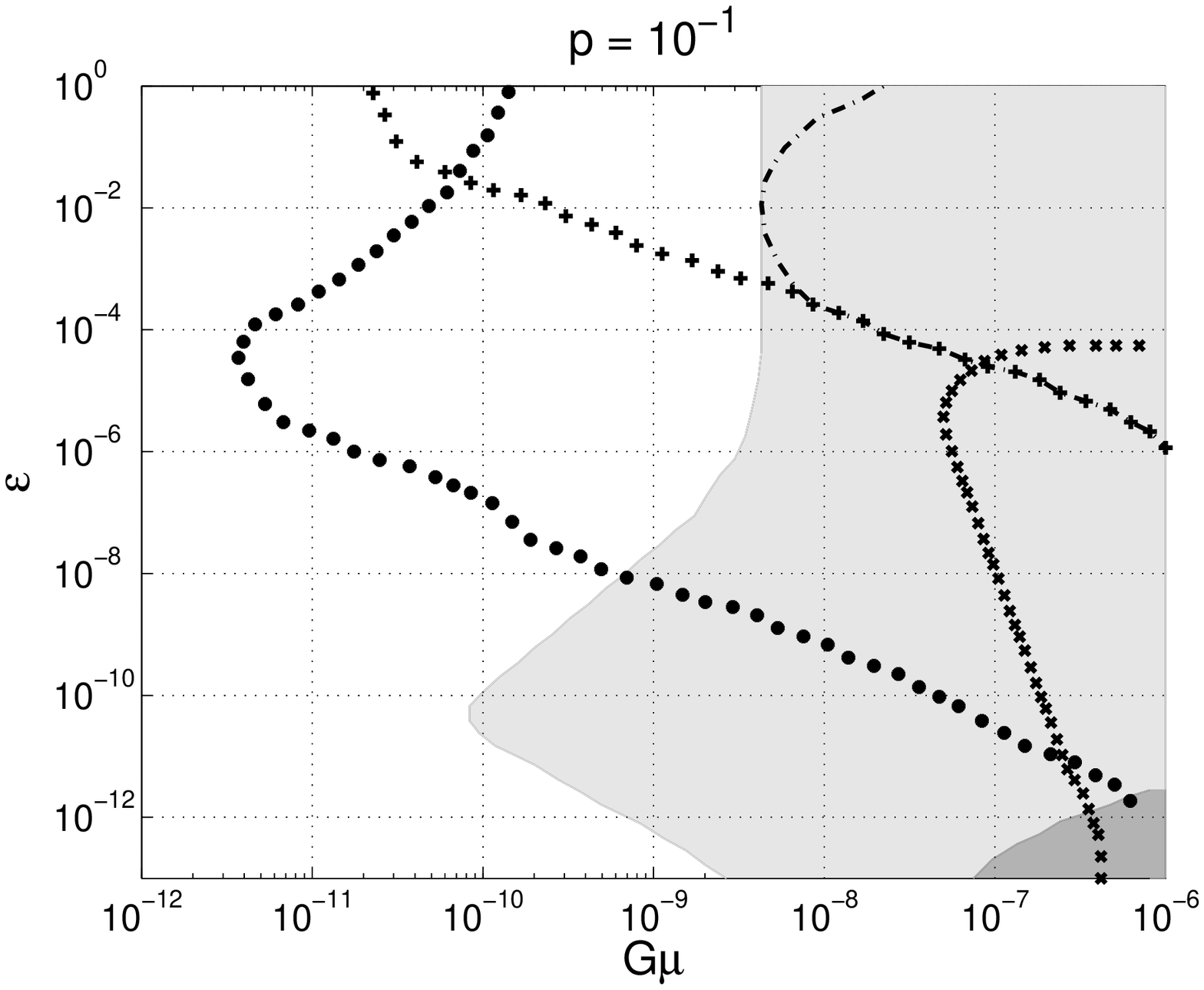} &
\hspace{1cm}
\includegraphics[width=2.8in]{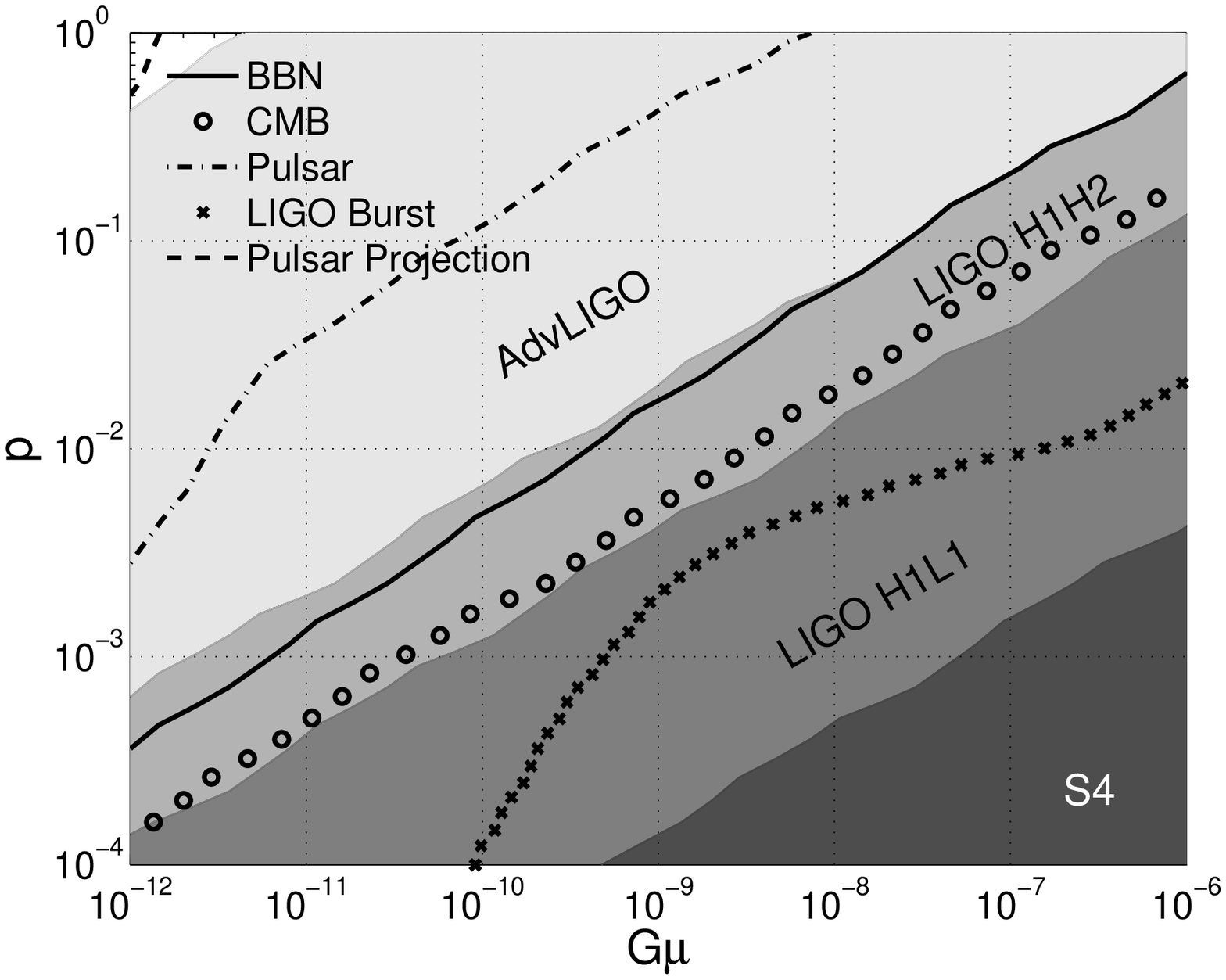} \\
\end{array}$
\caption{Top-left: Accessible regions in the $\varepsilon-G\mu$ plane
  for $p = 10^{-3}$ when loop sizes are determined by gravitational
  back-reaction. From darkest to lightest, they are: LIGO S4 limit,
  LIGO H1L1 projected sensitivity (cross-correlating the data from the
  4-km LIGO interferometers at Hanford, WA (H1) and Livingston, LA
  (L1)), LIGO H1H2 projected sensitivity (cross-correlating the data
  from the two LIGO interferometers at Hanford, WA (H1 and H2)), and
  AdvLIGO H1H2 projected sensitivity.  All projections assume 1 year
  of exposure and either LIGO design sensitivity or Advanced LIGO
  sensitivity tuned for binary neutron star inspiral search.  The
  solid black curve corresponds to the BBN bound, the dot-dashed curve
  to the pulsar bound, the $+$s to the projected pulsar sensitivity,
  the empty circles to the bound based on the CMB and matter spectra,
  the $\times$s to the projected sensitivity of the LIGO burst search,
  and the filled circles to the LISA projected sensitivity (accessible
  regions are to the right of the corresponding curves).  Top-right:
  Same as above for $p=10^{-2}$.  Bottom-left: Same as above for
  $p=10^{-1}$.  Bottom-right: Accessible regions in the $p-G\mu$ plane
  for the large long-lived loop models. The accessible regions are to
  the right of the corresponding curves. All models are within reach
  of LISA, and most are within the projected pulsar bound.}
\label{results}
\end{figure*}

If loop sizes at formation are determined by gravitational
back-reaction, then to a good approximation all loops have the same
length at formation and are short lived. We can take the loop
distribution to be $n(l,t)=(p\Gamma G \mu)^{-1}\delta(l-\alpha t)$
\cite{DV0}, where $\Gamma \sim 50$ is a constant related to the power
emitted by loops into GWs, and the cosmic time is a function of the
redshift, $t=t(z)$.  We parametrise the loop length using
$\varepsilon$ \cite{DV2}, taking $\alpha = \varepsilon \Gamma G \mu$.
In this case, the integral over lengths in Eq.~(\ref{e:Omega(f)2}) can
be replaced with Eq.~(59) of \cite{SCMMCR}, enhancing the string
density in the radiation era by a factor of $10$ \cite{DV0}.

However, recent simulations \cite{recentsims1,recentsims2}, suggest
that loop sizes at formation are related to network dynamics. In this case
loops may be large and long-lived, the loop distribution $n(l,t)$ is
more complicated (see Eqs.~(68-70) of \cite{SCMMCR}), and the integral
over lengths must be computed explicitly.

Damour and Vilenkin \cite{DV0} made the crucial observation that the
stochastic ensemble of GWs generated by a network of cosmic strings
includes large infrequent bursts, and that the computation of
$\Omega_{\rm gw}(f)$ should not be biased by including these large
rare events. When loops are small, all loops at a certain redshift are
the same size and produce the same amplitude events.  Hence, a cutoff
can be placed in the integral over redshifts to remove large events
for which the rate is smaller than the relevant time-scale of the
experiment (see Eq.~(6.17) of \cite{DV0}).  When loops are large the
situation is more complicated because at any given redshift there are
loops of many different sizes.  To deal with this problem, we use
Eqs.~(\ref{eq:hfz1}) and (\ref{eq:rate11}) and evaluate $dR/dzdh$, the
rate from cusps in redshift interval $dz$ and with strain in the
interval $dh$.  We then find the strain $h_*$ for which
\begin{equation}
R (>h_*) = \int _{h_*} ^\infty dh \frac {dR}{dh} = f.
\label{eq:rate14}
\end{equation}
Then, rather than Eq.~(\ref{e:Omega(f)2}) we evaluate
\begin{equation}
\Omega_{\rm gw}(f) = \frac{4 \pi^2}{3H^2_0}f^3 
\int_0 ^{h_*} dh \, h^2 \int dz \,  \frac {dR}{dzdh}.
\label{e:Omega(f)3}
\end{equation}
This procedure removes large amplitude events (those with strain
$h>h_*$) that occur at a rate smaller than $f$.

\noindent \textit{III. Results and discussion.} -- Our results take
the form of sections of cosmic string model parameter space either
constrained or allowed by past and future experiments (see
Fig.~\ref{results}). For simplicity we set $g_1=g_2=1$.

When loop sizes are given by gravitational back-reaction, we scan the
parameter space of re-connection probability ($10^{-3} < p < 1$),
dimensionless string tension ($10^{-12} < G\mu < 10^{-6}$), and the
size of the small loops ($10^{-13} < \varepsilon < 1$). For each point
in this parameter space, we calculate $\Omega_{\rm gw}(f)$.  Since the
most recent LIGO result \cite{S4stoch} was optimized for the frequency
independent spectrum, we first appropriately scale the observed LIGO
spectrum and variance, in order to optimize the search for the
calculated $\Omega_{\rm gw}(f)$ \cite{SMC}.  We perform similar
optimizations for the future projected sensitivities of LIGO and of
Advanced LIGO. For pulsar experiments (and the LISA sensitivity), we
exclude a model if it predicts a larger amplitude than the limit (or
projected sensitivity) at any frequency.  To compare a model with the
BBN bound, we perform the redshift integral in Eq. (\ref{e:Omega(f)2})
over redshifts $z \gtrsim 5.5 \times 10^9$.  Similarly, for the bound
based on the CMB and matter spectra, we integrate over $z \gtrsim
1100$.  Figure \ref{results} shows the accessible regions
corresponding to the different experiments and bounds. Several
conclusions can be inferred.  First, cosmic superstrings (with small
values of $p$) are more accessible because the spectrum amplitude is
inversely proportional to $p$ through its dependence on the loop
density. Second, there is much complementarity between different
experiments. The LIGO stochastic search is constraining models with
large $G\mu$ and small $\varepsilon$. Since the pulsar limit applies
at low frequencies ($f \approx 10^{-8}$ Hz), and due to the Heaviside
function on $\theta_m$ in the rate, the pulsar limit is more
constraining to models with larger loop lengths, i.e. large values of
$\varepsilon$ and $G\mu$. A similar argument applies to the LISA
projection, which is most sensitive around 1 mHz.  The LIGO burst
search is most sensitive to small $z$ and large strain cusps, which
also implies large $\varepsilon$ and large $G\mu$.  The BBN and CMB
bounds are not very sensitive to $\varepsilon$, because in the
large-$z$ limit $\Omega_{\rm gw} \sim G\mu/p$, i.e.  independent of
$\varepsilon$.  Third, the most recent LIGO stochastic bound has
already surpassed the BBN bound in an (admittedly small) part of the
parameter space.  This is because for some models a significant part
of the signal is produced after BBN.  Existing experiments and
indirect bounds already exclude a substantial part of the cosmic
string parameter space.  Future LIGO and LISA measurements will
continue to explore this parameter space.  Finally, although the LIGO
stochastic and LIGO burst searches are complementary, they also
overlap for large $G\mu$.  Hence, in the case of detection, the two
LIGO searches could potentially confirm each other.

To analyse the case when cosmic string loops are large at formation,
we take the loop distribution given by Eqs.~(68-70) of \cite{SCMMCR}
with the size of loops at formation given by $\alpha=0.1$
\cite{recentsims1}, and enhance number density of loops by a factor of
$1/p$.  We scan the parameter space given by $10^{-4} < p < 1$ and
$10^{-12} < G\mu < 10^{-6}$. Our estimate of the GW background in
these models is significantly larger than that of the small loop
models.  Hence, the current and future proposed experiments explore a
correspondingly larger part of the parameter space, as shown in the
bottom right panel of Fig.  \ref{results}.  In particular, values of
$p>0.1$ become more accessible.  Our results for current pulsar timing
experiments are substantially less constraining than the estimates of
Hogan \cite{hogan}, which relied on a less conservative pulsar timing
bound \cite{lommen}, and did not include effects of late-time
acceleration.  Currently, the pulsar limit is the most constraining,
but Advanced LIGO, LISA, and future pulsar timing experiments are
expected to explore all of this parameter space.  The BBN and CMB
bounds are consistent with, but somewhat weaker than, the pulsar
bound. For these models the constraints on superstrings from pulsar
timing experiments are particularly interesting. Notice the bound
rules out cosmic superstring models with $G\mu \gtrsim 10^{-12}$ when
the reconnection probability is $p \sim 10^{-3}$. Even for $p \sim
10^{-1}$ superstring tensions with $G\mu \gtrsim 10^{-10}$ are ruled
out.  Field theoretic strings and superstrings with $p \sim 1$ are
ruled out for $G\mu \gtrsim 10^{-8}$.

\noindent \textit{Acknowledgments.} -- We would like to thank the LSC
for making this study possible. We are indebted to Bruce Allen and
Irit Maor for numerous enlightening discussions.  We are further
grateful to Irit Maor for the use of her code to compute the
cosmological functions. We would also like to thank Alan Weinstein and
Richard O'Shaughnessey for carefully reading the paper and offering
clarifying comments and suggestions. The work of XS and JC was
supported by NSF grants PHY 0200852 and PHY 0421416.  The LIGO
Observatories were constructed by the California Institute of
Technology and Massachusetts Institute of Technology with funding from
the National Science Foundation under cooperative agreement
PHY-9210038. The LIGO Laboratory operates under cooperative agreement
PHY-0107417. LIGO Document Number LIGO-P060042-00-Z.

\end{document}